# Room-Temperature Polarity Control of the Anomalous Nernst Effect through sublattice-selective composition tuning in quasi-two-dimensional MnAlGe topological magnet

Nanhe Kumar Gupta[a, *], Keisuke Masuda[a], Masaaki Kakoki,[c],Benugopal Bairagya [a,b], Satoki Tazawa[a,b], Weinan Zhou[a] Hirofumi Suto[a], Ryo Toyama [a], Akio Kimura[c,d], Yuya Sakuraba [a,b,*]

[a] Research Center for Magnetic and Spintronic Materials (CMSM), National Institute for Materials Science (NIMS), Tsukuba, Japan

[b]Graduate School of Science and Technology, University of Tsukuba, Ibaraki, Japan.

[c]Graduate School of Advanced Science and Engineering, Hiroshima University, Japan.

[d]International Institute for Sustainability with Knotted Chiral Meta Matter (WPI-SKCM$^2$), Hiroshima University, Japan

E-mail: *GUPTA.Nanhekumar@nims.go.jp, SAKURABA.Yuya@nims.go.jp

Controlling the polarity of anomalous Nernst thermopower is a promising strategy for enhancing the performances of thermoelectric applications. However, realizing such control at room temperature (RT) in topological ferromagnets with high magnetic anisotropy ($K_{\mathrm{u}}$) remains challenging. Here, we report RT polarity control of the anomalous Nernst effect (ANE) in quasi-two-dimensional nodal-line MnAlGe epitaxial thin films through Al/Ge compositional tuning while preserving robust high $K_{\mathrm{u}}$. This polarity reversal originates from intrinsic Berry curvature contributions modulated by sublattice-selective carrier doping, as supported by spin-resolved electronic band structure analysis and hard X-ray photoemission spectroscopy. To demonstrate the practical feasibility, we also fabricated a meander-structured device combining MnAlGe with positive and negative polarity enhanced the thermoelectric output. Our results demonstrate that tuning the Fermi level relative to the nodal-line electronic structure while preserving high $K_u$ enables controllable ANE polarity reversal within a single material, providing a route toward RT transverse thermoelectric devices.

**Introduction.**

Efficient conversion of waste heat into electricity is vital for powering small-scale electronic and IoT devices[1]. Among thermoelectric energy-conversion mechanisms, the anomalous Nernst effect (ANE) in magnetic materials has attracted growing attention owing to its simple planar device geometry, in contrast to conventional Seebeck-effect (SE) modules[2]. The inherent simplicity of ANE-based devices enables easy integration, scalability, and cost-efficiency, making them attractive for thermopile applications such as a heat flux sensor (HFS) [3] and thermoelectric power generator [4,5]. For practical ANE-based devices, it is essential to develop materials that exhibit both positive and negative anomalous Nernst coefficients ( $\mathrm{S_{ANE}}$ ) together with high magnetic anisotropy ( $K_\mathrm{u}$ ) at room temperature (RT), since large magnetic anisotropy ensures stable magnetization and reliable operation under ambient conditions [6]. Although several materials that exhibit either large $\mathrm{S_{ANE}}$ [7,8,9] or high $K_\mathrm{u}$ [10,11] at room temperature have been explored, most of the materials with large $\mathrm{S_{ANE}}$ often have small $K_\mathrm{u}$, and vice versa, making their coexistence in a single system a major challenge [2]. In fact, materials that exhibit controllable bipolar ANE together with high $K_\mathrm{u}$ at RT remain elusive. Such bipolarity within a single materials platform would eliminate the need for integrating different positive- and negative-ANE materials, greatly simplifying the fabrication of high-performance ANE thermopiles.

Fundamentally, such bipolarity can, in principle, be achieved in topological magnets, where tuning the Fermi level ($E_\mathrm{F}$) across the exchange gap reverses the sign of the ANE [12]. Indeed, Weyl and nodal- line semimetals exhibit large ANE signals arising from enhanced Berry curvature near singular band features. However, controlled polarity reversal in practical materials remains challenging because multiple Fermi-surface contributions often obscure the intrinsic Berry-curvature-driven response, while chemical substitution used for $E_\mathrm{F}$ tuning tends to distort the underlying topological band structure itself. Recent studies have demonstrated ANE polarity reversal in systems such as $Co_3Sn_2S_2$-based Weyl magnets (via Ni or In doping) [12].Although this work demonstrates the feasibility of polarity control, the finite ANE appears only in low temperature region because of their low Curie temperature (~180 K). Realizing Berry-curvature-driven ANE polarity control in a single-phase magnetic material combining large $K_\mathrm{u}$ and RT stability remains an open challenge.

Layered - and quasi-two-dimensional topological ferromagnets (FM) are promising material platforms to realize both high $K_u$ and polarity control of ANE because of their low dimensionality-induced intrinsic strong magnetic anisotropy and higher controllability of $E_F$ with preserving the topological band features due to the reduced Fermi surface complexity [13,15]. This approach could provide a pathway to achieve bipolar ANE within a single-phase material, while maintaining strong magnetic anisotropy. In this respect, quasi-two-dimensional material MnAlGe emerges as an excellent candidate. In MnAlGe, Mn layers are separated by nonmagnetic Al and Ge layers, leading to two dimensionality-induced high $K_u$[13,14]. In addition, recent studies reported large Berry curvature originating from simple two-dimensional topological nodal line [13,15], making it promising material to tune the polarity of ANE[16]. In this study, we propose to tune the composition ratio of Al/Ge in MnAlGe as the simplest method to control the $E_F$ while preserving the underlying topological band features and maintaining high $K_u$.

We fabricate the composition-spread epitaxial $Mn(Al_{1-x}Ge_x)_2$ thin film ($0 \leq x \leq 1$) and demonstrate clear RT polarity reversal of the ANE, achieved solely by tuning the Al/Ge ratio. Remarkably, this polarity reversal occurs without disrupting the C38-type crystal structure or robust magnetic anisotropy, indicating that the underlying topological features remain intact. The observed behavior is attributed to Berry-curvature-driven intrinsic contributions modulated by Fermi level shifting, as supported by theoretical analysis of electronic structures and hard X-ray photoemission spectroscopy (HAXPES). Beyond these fundamental insights, we fabricated a poly-crystalline MnAlGe based thermopile module on $SiO_2$ substrates and demonstrate that the polarity reversal enhances the output ANE voltage together with the strong in-plane magnetic anisotropy at the device level. These results highlight the realization of polarity reversal of anomalous Nernst thermopower in a high $K_u$ quasi two-dimensional topological FM at RT, marking a crucial step toward next generation ANE-based thermoelectric and sensing applications.

## Results and Discussion

To illustrate the effect of Al/Ge substitution on the crystal structure, Fig. 1a shows the schematic unit cells of $Mn(Al_{1-x}Ge_x)_2$ for representative compositions $x = 0.3$, 0.5, and 0.7. In all cases, Ge substitutes selectively at the Al sublattice, while the Mn sublattice and overall $Cu_2Sb$-type (C38) crystal structure remain unchanged. To clarify the effect of Al:Ge composition on the electronic structure, the spin-resolved DOS of $Mn(Al_{1-x}Ge_x)_2$ for $x$=0.3 to 0.7 is shown in Fig. 1b, where $E_F$ denotes the Fermi level. With increasing Ge concentration, the exchange splitting energy between majority-spin and minority-spin DOS gradually decreases. Notably, focusing only on the minority-spin DOS near the $E_F$, it signifies that an increase in Ge concentration $x$ works to produce a rigid-band-like electron doping effect. The magnitude of the shifting of the $E_F$ with respect to $x$ = 0.5 is about + 110 meV for $x$ =0.3 and – 157 meV for $x$ = 0.7. Thus, the total shift from the $E_F$ $x$ =0.3 to 0.7 is about 267 meV. Large minority-spin DOS near the $E_F$ dominantly originates from localized $d$-bands of Mn which was predicted to form topological nodal points [13]. Guin *et al.* showed that, in the absence of spin-orbit coupling, MnAlGe hosts symmetry-protected nodal lines in the minority-spin channel arising from mirror symmetries. The corresponding minority-spin band structures, calculated without spin-orbit coupling are shown in Fig. 1c for representative compositions from $x$ = 0.3 to 0.7. We first reproduce these nodal-line features for stoichiometric MnAlGe ($x$ = 0.5), validating our computational approach. With Al/Ge composition ratio tuning, the symmetry-protected nodal-line surrounded by red circles is preserved over a wide range of compositions with a slight shifting of the wave number of nodal points in the X- Γ- R symmetry line. With increasing (decreasing) Ge content, electron (hole) doping shifts the $E_F$ upward (downward), leading to a systematic movement of the nodal features toward the $E_F$. This selective tunability is enabled by the layered C38 structure, in which the magnetic Mn layers are spatially separated from the nominally nonmagnetic Al/Ge layers. Varying the Al/Ge ratio therefore provides sublattice-selective carrier doping that shifts the $E_F$ while the Mn-$d$-derived bands remain largely unchanged. Consequently, the topological nodal-line electronic structure is preserved over $0.3 \leq x \leq 0.7$, whereas $E_F$ is shifted by approximately 310 meV. This value is consistent with the shift estimated from the minority-spin DOS in Fig. 1b. This sublattice-selective tuning provides a direct route to modulate the Berry curvature near $E_F$ without substantially perturbing the magnetic Mn-derived electronic structure.

To experimentally realize this $E_F$-tuning strategy, we fabricated composition-spread epitaxial $Mn(Al_{1-x}Ge_x)_2$ thin films on MgO(001) substrates using the combinatorial sputtering technique that have been well established in earlier studies[17–20]. Structural characterization by x-ray diffractometer clearly reveals that the $Cu_2Sb$-type C38 single phase is stabilized over a wide composition range of ($0.35 \leq x \leq 0.70$) (Supplementary Fig. S2), whereas Al-rich ($0.01 < x \leq 0.21$) and Ge-rich ($0.70 < x \leq 0.97$) compositions show phase evolution and secondary-phase formation.

We therefore focus on the single-phase C38 region for the subsequent electronic and transport analyses.

The $E_F$-tuning effect is experimentally examined by hard X-ray photoelectron spectroscopy (HAXPES). We measured the $x$ dependence of Mn $2p$, Al $1s$, and Ge $2p$ core-level spectra of $Mn(Al_{1-x}Ge_x)_2$ for $0.2 \leq x \leq 0.83$. Figure 2a–c show the core-level spectra, and the corresponding binding-energy evolution is summarized in Fig. 2d–f. Details of the spectral analysis are provided in the Supplementary Material (Figs. S6). Within the single-phase C38 region ($0.35 \leq x \leq 0.70$), all three core levels show an overall shift toward higher binding energy (i.e., lower $E$-$E_F$ ), with increasing Ge concentration. The common direction of the core-level shifts provides spectroscopic evidence for an upward shift of the chemical potential associated with electron doping by Ge substitution, consistent with the theoretical calculations in Fig. 1b. In particular, the Ge $2p$ core level shifts by approximately 0.1–0.15 eV over the relevant composition range, which is comparable to the calculated chemical-potential shift of approximately 0.14 eV between $x = 0.41$ and 0.65. This semi-quantitative agreement supports the predicted $E_F$-tuning induced by Al/Ge compositional control. The magnitude of the shift, however, is clearly element dependent. In particular, the Al $1s$ and Mn $2p$ core levels exhibit larger and more complex variations than Ge $2p$. Core-level binding energies generally contain not only the chemical-potential shift but also site-dependent contributions associated with changes in the local charge distribution, electrostatic potential, and final-state screening [21,22]. The deviations of the Al $1s$ and Mn $2p$ shifts from the calculated $E_F$-shift are therefore attributed to such local chemical-environment effects induced by Al/Ge substitution. Thus, although the absolute shift of an individual core level cannot be directly equated with the chemical-potential shift, the common binding-energy trend and, particularly, the energy scale of the Ge $2p$ shift provide experimental support for the theoretically predicted electron-doping-induced $E_F$ shift.

Figures 3a -b show the $H$-dependent AHE and ANE signals, respectively. For compositions $0.35 \leq x \leq 0.69$, both AHE and ANE signals exhibit square-shaped hysteresis loops with sharp switching, indicating strong perpendicular magnetic anisotropy (PMA). The presence of strong PMA is further supported by the in-plane-field AHE measurement, which shows a large effective in-plane saturation field, $(\mu_0 H_k^{\mathrm{eff}})$. This field is substantially larger than the saturation field in the out-of-plane configuration, confirming that the in-plane direction is the magnetic hard axis [see

SM Fig.S8]. Importantly, $\mu_0 H_k^{\mathrm{eff}}$ remains large and nearly composition-independent, demonstrating robust magnetic anisotropy throughout the C38 phase. The $K_u$ (Fig. 3c),which was evaluated by measuring the saturation magnetization in the separately fabricated uniform composition films with the $x$ = 0.32-0.7 (shown in SM), remains high (~ 0.40 - 0.55 $\mathrm{MJm^{-3}}$) and are comparable to those previously reported for bulk MnAlGe [13]. The persistence of large $K_u$ across Ge concentration in MnAlGe is qualitatively consistent with theoretical reports that the magnetocrystalline anisotropy is primarily governed by spin–orbit-coupled Mn 3$d$ states in the tetragonal C38 structure [23,14] , making it largely insensitive to the Al/Ge composition ratio. This robustness enables that Al/Ge compositional tuning for carrier/transport response while preserving the Mn-derived electronic structure responsible for the PMA.

Despite nearly identical coercive fields and similar hysteresis shapes, AHE and ANE show distinctly different transport behavior. The most striking observations are that ANE shows polarity change around $x \approx 0.41$, while the AHE does not. The extracted anomalous Hall resistivity ($\rho_{yx}^{A}$) values are summarized with respect to $x$ in Fig. 3d. $\rho_{yx}^{A}$ gradually increases with $x$ and takes peak ~9.1 μΩ·cm at $x$ = 0.46 and then decreases with $x$ to 2.2 μΩ·cm. The observed value ~7 μΩ·cm at nearly stochiometric MnAlGe ($x$ = 0.52) is almost comparable to values reported for single crystals and epitaxial thin films of MnAlGe [13,16]. Additional transport parameters, including the anomalous Hall angle and longitudinal resistivity, were also extracted and are shown in Fig. S7. Figure 3e shows the Seebeck coefficient ($S_{\mathrm{SE}}$) as a function of Ge concentration. At $x$ = 0.36-0.41, $S_{\mathrm{SE}}$ remains relatively small (~5 $\mu\mathrm{V\,K^{-1}}$). A sharp enhancement occurs at $x$ = 0.46, where $S_{\mathrm{SE}}$ reaches ~11 $\mu\mathrm{V\,K^{-1}}$, followed by a gradual increase with higher Ge concentration. The $S_{SE}$ value observed near $x$ ≈0.52 also agrees well with previously reported values (~ 10–11 $\mu\mathrm{V\,K^{-1}}$) for stoichiometric MnAlGe thin films[16,24]. Furthermore, $S_{\mathrm{SE}}$ remains positive throughout the entire composition range, confirming the dominance of hole-like ($p$-type) carriers in MnAlGe.

The $S_{\mathrm{ANE}}$ values exhibit a sharp increase and sign reversal from positive (+0.09 $\mu\mathrm{V\,K^{-1}}$) at $x$ = 0.41 and to negative (-0.63 $\mu\mathrm{V\,K^{-1}}$) at $x$ = 0.46 as shown in Fig. 3f. Such sharp change with the sign reversal between $x$ = 0.41 and $x$ = 0.46 (corresponding to 7-fold change in magnitude) is in contrast to the change of $\rho_{\mathrm{yx}}^{\mathrm{A}}$ (1.4-fold) and $S_{SE}$ (1.8-fold) without no sign reversal. The $S_{\mathrm{ANE}}$ can be expressed as $S_{\mathrm{ANE}} = \rho_{xx}\, \alpha_{xy}^{\mathrm{A}} - \rho_{yx}^{\mathrm{A}} \alpha_{xx}$. The first term, $\rho_{xx}\, \alpha_{xy}^{\mathrm{A}}$ defined as $S_{\mathrm{I}}$, the second term, $\rho_{yx}^{A}\alpha_{xx}$ defined as $S_{\mathrm{II}}$. This can also be written as $S_{\mathrm{II}} = -S_{\mathrm{SE}}\,\theta_{\mathrm{AHE}}$, where $\theta_{\mathrm{AHE}}(= \rho_{\mathrm{yx}}^{\mathrm{A}}/\rho_{xx})$ is the anomalous Hall angle respectively[25]. As summarized in Fig. 3f, $S_{\mathrm{I}}$ and $S_{\mathrm{II}}$ exhibit distinct dependence on $x$. It is clearly observed that the sign of $S_{\mathrm{II}}$ term is negative in whole composition region, thus the polarity reversal of $S_{\mathrm{ANE}}$ primarily governed by the $S_I$ term. Since $\rho_{xx}$ cannot be negative, the sign of $S_{\mathrm{ANE}}$ directly reflects the behavior of the $\alpha_{xy}^{\mathrm{A}}$.

To further elucidate the origin of the anomalous transport, considering the electron doping effect by Ge substitution (Fig.1b) and its experimental verification by HAXPES(Fig.2)), here we compare the experimentally evaluated $x$ dependence of $\sigma_{xy}^{A}$ $(=\frac{\rho_{yx}^{A}}{{\rho_{yx}^{A}}^{2}+\rho_{xx}^{2}})$ and $\alpha_{xy}^{A}$ with first-principles calculations of the chemical potential $\mu$ dependence of the $\sigma_{xy}^{\mathrm{int}}$ and $\alpha_{xy}^{\mathrm{int}}$ for the stochiometric MnAlGe ($x$ = 0.5)[24], as summarized in Figs. 4a-d. A comparison between experiment and theory reveals that clear mismatch the position of $E_{\mathrm{F}}$. In the calculations, $\sigma_{xy}^{A}$ exhibits a maximum in the vicinity of $E_{\mathrm{F}}$ for $x$ = 0.5, labeled as α′, whereas experimentally the peak appears at $x \approx 0.41$, labeled as $\alpha$. By aligning the experimental $\alpha$ peak with the calculated $\alpha'$ peak, the required shift is estimated to be approximately +200 meV in the stochiometric MnAlGe region ($x$ = 0.5) relative to the calculated electronic structure. Notably, slight electron-doping effect was also reported in stoichiometric bulk single crystalline MnAlGe [13].

Considering this electron-doping in $x$ = 0.5, the experimental $x$ dependence of both $\sigma_{xy}^{A}$ and $\alpha_{xy}^{A}$ qualitatively well matches with theoretical energy dependence. The most notable feature is that $\alpha_{xy}^{\mathrm{A}}$ becomes negative at around $x$ = 0.5 (region $b$ in Fig.4c). Considering the $E_{\mathrm{F}}$ shift of approximately +200 meV, the $x$ = 0.5 composition region can be assigned to the $b'$ region in the energy dependence of the intrinsic $\alpha_{\mathrm{xy}}^{\mathrm{int}}$ shown in Fig. 4d. Within this framework, the experimentally observed sign reversal of $\alpha_{\mathrm{xy}}^{\mathrm{A}}$ from negative to positive upon both hole doping (decreasing $x$) and electron doping (increasing $x$) can be consistently explained by the correspondence between the $a$ and $a'$ regions and between the $c$ and $c'$ regions, respectively. The energy separation between the $a'$ and $c'$ regions (edges of positive $\alpha_{xy}^{\mathrm{int}}$ regions) is estimated to be approximately 150 meV. This value is in almost same energy range of the predicted $E_{\mathrm{F}}$ shift of about 267 meV between $x = 0.3$ and $0.7$, and 138 meV between $x = 0.41$ and $0.65$estimated from Fig. 1b. The latter value was estimated by linear interpolation between the calculated $E_F$ shifts for $x = 0.3$ and $0.7$ .+, This agreement further supports the proposed qualitative correspondence between composition and energy shift. A similar trend is also observed for $\sigma_{xy}^{\mathrm{A}}$. Experimentally, $\sigma_{xy}^{\mathrm{A}}$exhibits a maximum at $x = 0.41$ on the hole-doped side, while a slight enhancement appears around $x = 0.65$ on the electron-doped side. These features suggest a correspondence between the $\alpha$ and $\alpha'$ regions and between the $\gamma$ and $\gamma'$ regions in the calculated energy dependence of $\sigma_{xy}^{\mathrm{int}}$. At first glance, an apparent discrepancy exists between experiment and theory because both $\sigma_{xy}^{\mathrm{A}}$ and $\alpha_{xy}^{\mathrm{A}}$ exhibit maxima at $x = 0.41$, whereas the calculated peaks of $\sigma_{xy}^{\mathrm{int}}$ and $\alpha_{\mathrm{xy}}^{\mathrm{int}}$ are located at $\mu$ = +20 meV and +80 meV, respectively. However, because the composition step ($\Delta x \approx 0.06$) in present experiment is considerably larger than the energy scale associated with the fine structures predicted by the theoretical calculation, the experimental $x$ dependence is unable to fully resolve such detailed features. Consequently, the true peak and dip positions may not be accurately captured in the experimental data, resulting in an overlap of the maxima in $\sigma_{xy}^{\mathrm{A}}$ and $\alpha_{xy}^{\mathrm{A}}$.

Consequently, we conclude that the shifting of the Fermi level by Al/Ge composition tuning effectively moves the nodal-line-derived Berry-curvature hotspots, enabling control over the sign of $\alpha_{xy}^{\mathrm{A}}$. It should be noted that, although the compositional trends agree well with theory, the experimental magnitudes of $\sigma_{xy}^{\mathrm{A}}$ and $\alpha_{xy}^{\mathrm{A}}$ are suppressed by nearly an order of magnitude. This suggests that, although the ANE in MnAlGe predominantly originates from an intrinsic Berry-curvature mechanism, additional effects reduce its magnitude, which will be discussed later.

Having established the intrinsic Berry-curvature origin of the ANE polarity reversal, we compare our approach with previously reported room-temperature ANE materials and strategies for polarity control. ANE polarity control has been demonstrated in $Co_xGd_{1-x}$[26], $Fe_3Ln$ (Ln = Ho, Tb, Gd, Er)[27], Co-substituted $Nd_2Fe_{14}B$[28], and Cr- doped MnAlGe[16],. However, these approaches rely on ferrimagnetic compensation, magnetic-element substitution, and or transition-metal doping. Interestingly, Cr-doped MnAlGe provides a closer comparison by realizing ANE polarity tuning within the MnAlGe family[16]. However, microscopic mechanism responsible for sign reversal and its relation to the underlying topological nodal-line electronic structure remains unclear. In contrast, the present $Mn(Al_{1-x}Ge_x)_2$ films realize ANE polarity reversal at RT solely through Al/Ge compositional tuning while preserving the high- $K_{\mathrm{u}}$ in C38-type MnAlGe phase, without introducing rare-earth elements, noble metals, or additional magnetic dopants. This sublattice-selective carrier-doping strategy enables direct Fermi-level control of Berry-curvature-driven transport while preserving the underlying topological nodal-line electronic structure. These results establish a materials platform in which high magnetic anisotropy, topological band features and tunable anomalous thermoelectric transport coexist, and highlight sublattice-selective carrier doping as a general strategy for engineering transverse thermoelectric responses.

**Device Demonstration**

Having established composition-controlled ANE polarity reversal in epitaxial $Mn(Al_{1-x}Ge_x)_2$ films, we next demonstrated this functionality into practical heat-flux-sensing geometry. Since ANE-based heat flux sensors operate under an out-of-plane $\nabla T$ with a stable in-plane remanent magnetization for zero-field voltage generation[3,4,5], we fabricated a poly-crystalline MnAlGe (MAG) on a $SiO_2$ substrate to obtain large coercivity to the in-plane direction. The ANE thermopiles were patterned with the MAG wires with positive ($x$ = 0.39) and negative ($x$ = 0.54) $S_{\mathrm{ANE}}$ which were aligned alternatively (see SM for details fabrication process). Figure 5a shows the schematic device lay out schematic, in this geometry, positive (light red) and negative (blue) $S_{\mathrm{ANE}}$ are connected in series. The device response clearly shows the thermopile operation, compared with individual positive and or negative $S_{\mathrm{ANE}}$, the output voltage increases systematically with the number of connected pairs, reaching significantly higher values at $n$ = 2 and 10 (Fig. 5b). In addition, although both MAG films with $x$ = 0.39 and 0.54 have a poly-crystalline nature, both compositions retain large coercivity of about 0.3 T and large remanent magnetization for in-plane magnetic fields, which is advantageous for the practical HFS application[3,4,5]. The persistence of opposite ANE polarities in the polycrystalline films further indicates that the composition-controlled sign reversal is robust against the loss of epitaxy. Although the magnitude of the ANE in MnAlGe is moderate, the present results demonstrate a materials-design concept in which topological two-dimensional layered magnets enable independent tuning of ANE polarity while retaining robust magnetic anisotropy. This strategy provides a promising route toward monolithic ANE thermopiles based on a single materials platform.

Despite the successful realization of polarity control in anomalous Nernst thermopower and device operation, the experimentally observed $\alpha_{\mathrm{xy}}^{\mathrm{A}}$ is substantially smaller than theoretical predictions, as shown in Fig.4. To understand the suppression of the experimental magnitudes of $\sigma_{xy}^A$ and $\alpha_{xy}^A$, we consider the role of spin fluctuations, which are expected to be enhanced in two-dimensional ferromagnets and are not captured in zero-temperature band-structure calculations [24,29,30]. In our previous studies on the epitaxial MnAlGe film, we detected a signature of strong thermally induced electron–magnon scattering through the longitudinal magnetoresistance (LMR) measurements, suggesting an remarkable spin-fluctuation in MnAlGe at finite temperature [24]. Direct experimental evidence for enhanced spin fluctuations is also obtained in present $Mn(Al_{1-x}Ge_x)_2$ composition spread film through the LMR measurements (see SM Fig. S9). Notably, this behavior persists across the entire composition range, suggesting that spin fluctuations are a robust feature over a wide range of Al/Ge compositions. Although spin fluctuations are predicted to strongly influence anomalous transport through magnon-mediated scattering[31,32], their role in modifying intrinsic or extrinsic contributions remained unresolved for MnAlGe. The fact that the compositional evolution of $\sigma_{xy}^A$ and $\alpha_{xy}^A$ are qualitatively well reproduced by the theoretical calculation suggests that spin fluctuations predominantly affect the magnitude of intrinsic Berry-curvature contribution rather than extrinsic contribution. These findings suggest that the remaining bottleneck for layered topological magnets is the suppression of finite-temperature spin fluctuations that substantially reduce the anomalous transport response. Future materials design should therefore focus on strengthening magnetic exchange interactions while preserving the topological electronic structure, thereby enabling the realization of the theoretically predicted ANE performance.

## Conclusion

In summary, we investigated the C38-type $Mn(Al_{1-x}Ge_x)_2$ system across a wide compositional range and demonstrated room-temperature polarity reversal of the ANE in a high-$K_u$ topological magnet through Al/Ge composition tuning. The polarity reversal originates from the distinct sensitivity of the ANE to Berry curvature near the Fermi level. The good agreement between experiment and theory confirms that both AHE and ANE are dominated by intrinsic Berry-curvature-driven mechanisms. Beyond the fundamental study, we further demonstrate a meander-type ANE thermopile in which combining positive- and negative-polarity MnAlGe enhances the transverse thermoelectric output. The persistence of opposite ANE polarities in both epitaxial and polycrystalline films highlights sublattice-selective carrier doping as a robust route to tune the Fermi level in quasi-two-dimensional layered magnets. These results establish controllable polarity reversal of anomalous Nernst thermopower through electronic-structure engineering as a promising strategy for designing room-temperature transverse thermoelectric functionalities in topological ferromagnets. In particular, the exploration of systems with reduced spin fluctuations could enable substantially enhanced $S_{ANE}$ and improved thermoelectric performance, providing a promising direction for next-generation transverse thermoelectric technologies.

**Materials and Methodology**

Epitaxial $Mn(Al_{1-x}Ge_x)_2$ thin films with a thickness of 30 nm ($x$ = 0 -1) were fabricated on MgO (001) substrates by combinatorial magnetron co-sputtering. Details of the deposition process and a schematic illustration (SM Fig. S1) are provided in the supplementary material information (SM). In addition, a separate set of uniform-composition films ($x$ = 0.32, 0.38, 0.4, 0.46, 0.48, 0.51, 0. 64 and 0.70) was grown on MgO (001) substrates to perform magnetization measurements for evaluating the magnetic anisotropy energy. The crystal structure was characterized by X-ray diffraction with CuKα radiation, with results summarized in Supplementary Fig. S2. Electrical and thermoelectric transport measurements were performed using a home-built holder stage capable of applying a controlled temperature gradient in a Versa lab (Quantum Design). Further details of the transport measurement procedures and device fabrication are described in the SM. A photograph of the thermoelectric measurement holder is shown in SM (see Supplementary Fig. S5). Theoretical calculations of the anomalous Hall and Nernst conductivities in MnAlGe were conducted based on the density-functional theory (DFT) including the spin-orbit interaction. Electronic structures of $Mn(Al_{1-x}Ge_x)_2$ were calculated using the Korringa-Kohn-Rostoker method combined with the DFT. Details of these calculation methods are provided in the SM. Spin-integrated hard X-ray photoemission spectroscopy (HAXPES) experiments were carried out at beamline BL09U of NanoTerasu details provided in SM. To demonstrate the device concept (see SM for fabrication details), two films with opposite ANE polarity ($x$= 0.39 and 0.54) were deposited on $SiO_2$ substrates. A clear enhancement of the ANE signal was observed with an increasing number of pairs in the meander structure.

**Acknowledgements**

The authors thank N. Kojima and T. Hiroto for technical support with the thermoelectric measurement, device fabrication and XRD measurement, respectively.

**Data availability**

The data that supports the findings of this study are available from the corresponding authors upon reasonable request.

**Competing interests**

The authors declare no competing interests.

**Author contributions**

N.K.G. and S.Y. conceived the study and designed the experiments. N.K.G. fabricated the samples and devices and performed the experimental characterization, with support from B.B. and W.Z. S.T., R.T. and H.S. contributed to scientific discussions and interpretation of the results. K.M. performed and analyzed the density functional theory calculations and contributed to the interpretation of the calculated anomalous transport properties. N.K.G. and S.Y. performed the HAXPES measurements. M.K. analyzed the HAXPES data, with A.K. supervising the HAXPES analysis and interpretation. N.K.G. and S.Y. analyzed and interpreted the experimental data and wrote the manuscript with input and comments from all authors. S.Y. coordinated and supervised the project.

**Funding**

This work was supported by JST ERATO "Magnetic Thermal Management Materials" (Grant No. JPMJER2201, JST CREST "Exploring Innovative Materials in Unknown Search Space" (Grant No. JPMJCR21O1) and A-STEP (Stage II, Full-scale type) (Grant No. JPMJTR253A) from JST, Japan. he Grants-in-Aid for Scientific Research (C) (KAKENHI; Grant No. 25K08463) from JSPS, Japan, the Grants-in-Aid for Scientific Research (B) (KAKENHI; Grant No. 26K00993) and Data Creation and Utilization-Type Material Research and Development Project(Digital Transformation Initiative Center for Magnetic Materials; Grant No. JPMXP1122715503). Theoretical calculations in this work were performed on the Numerical Materials Simulator at NIMS.

**ORCID**

Nanhe Kumar Gupta**:** https://orcid.org/0000-0002-5488-4986

Keisuke Masuda: https://orcid.org/0000-0002-6884-6390

Masaaki Kakoki: https://orcid.org/0009-0008-5096-498X

Weinan Zhou: https://orcid.org/0000-0003-2946-9913

Ryo Toyama: https://orcid.org/0000-0002-7398-5803

Benugopal Bairagya: https://orcid.org/0009-0006-4549-8908

Hirofumi Suto: https://orcid.org/0000-0003-4387-5862

Akio Kimura: https://orcid.org/0000-0002-1501-3918

Yuya Sakuraba: http://orcid.org/0000-0003-4618-9550

## REFERNCES

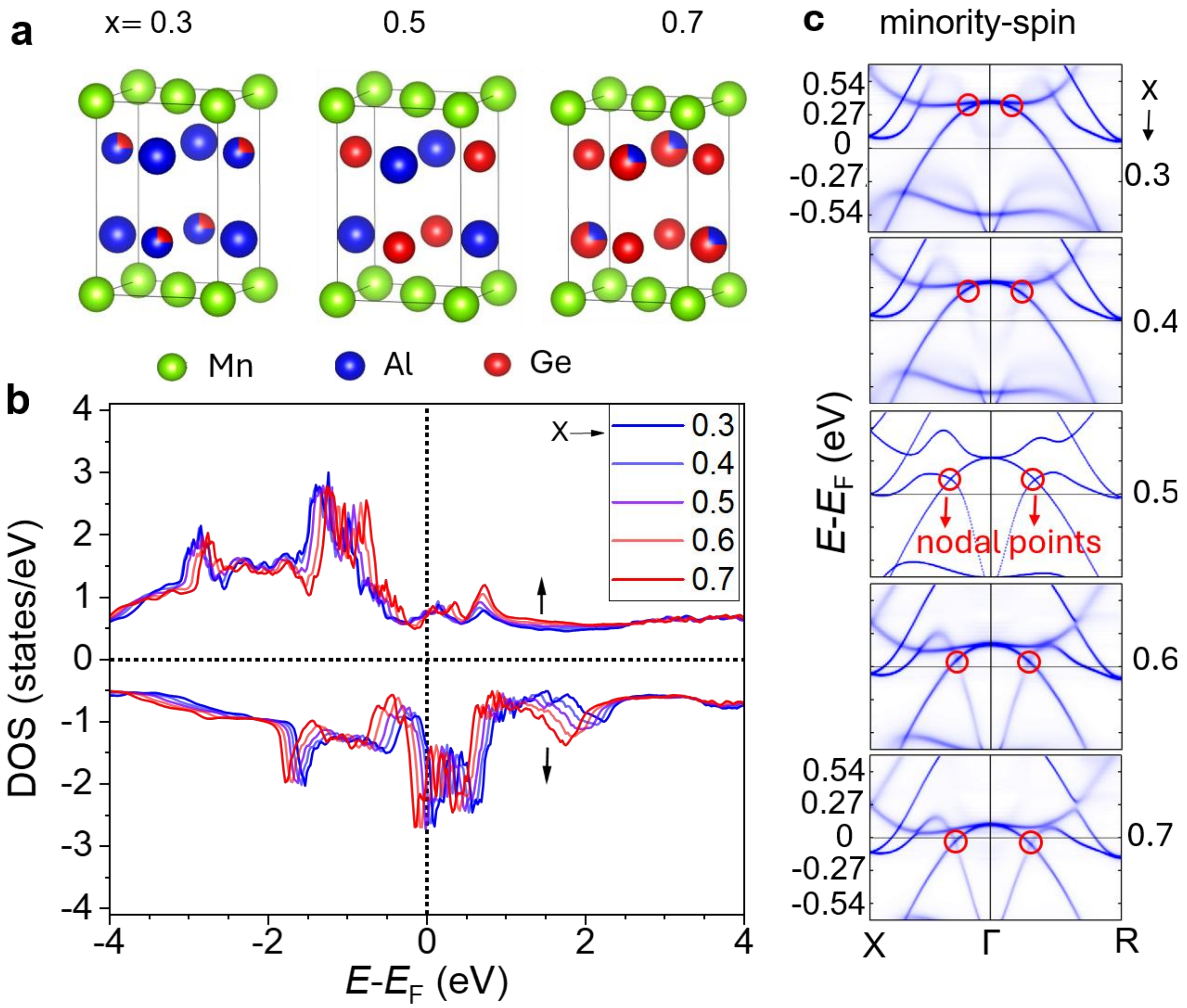

**Fig. 1**. a, Crystal structures of $Mn(Al_{1-x}Ge_x)_2$ for representative compositions $x = 0.3$, 0.5, and 0.7, illustrating the evolution of atomic occupation with Ge substitution. Green, blue, and red spheres denote Mn, Al, and Ge atoms, respectively. b, Composition dependence of the spin-resolved density of states for $Mn(Al_{1-x}Ge_x)_2$. $0.3 \leq x \leq 0.7$, plotted relative to the Fermi level ($E_F = 0$). c, Spin-resolved electronic band structures (Spin-down) calculated without spin-orbit coupling for representative compositions. The circles highlight nodal-line features at the X- Γ-R symmetry line. In c, all band-structure panels share the same energy scale, with $E_F$ set to zero; repeated y-axis tick labels are omitted for clarity.

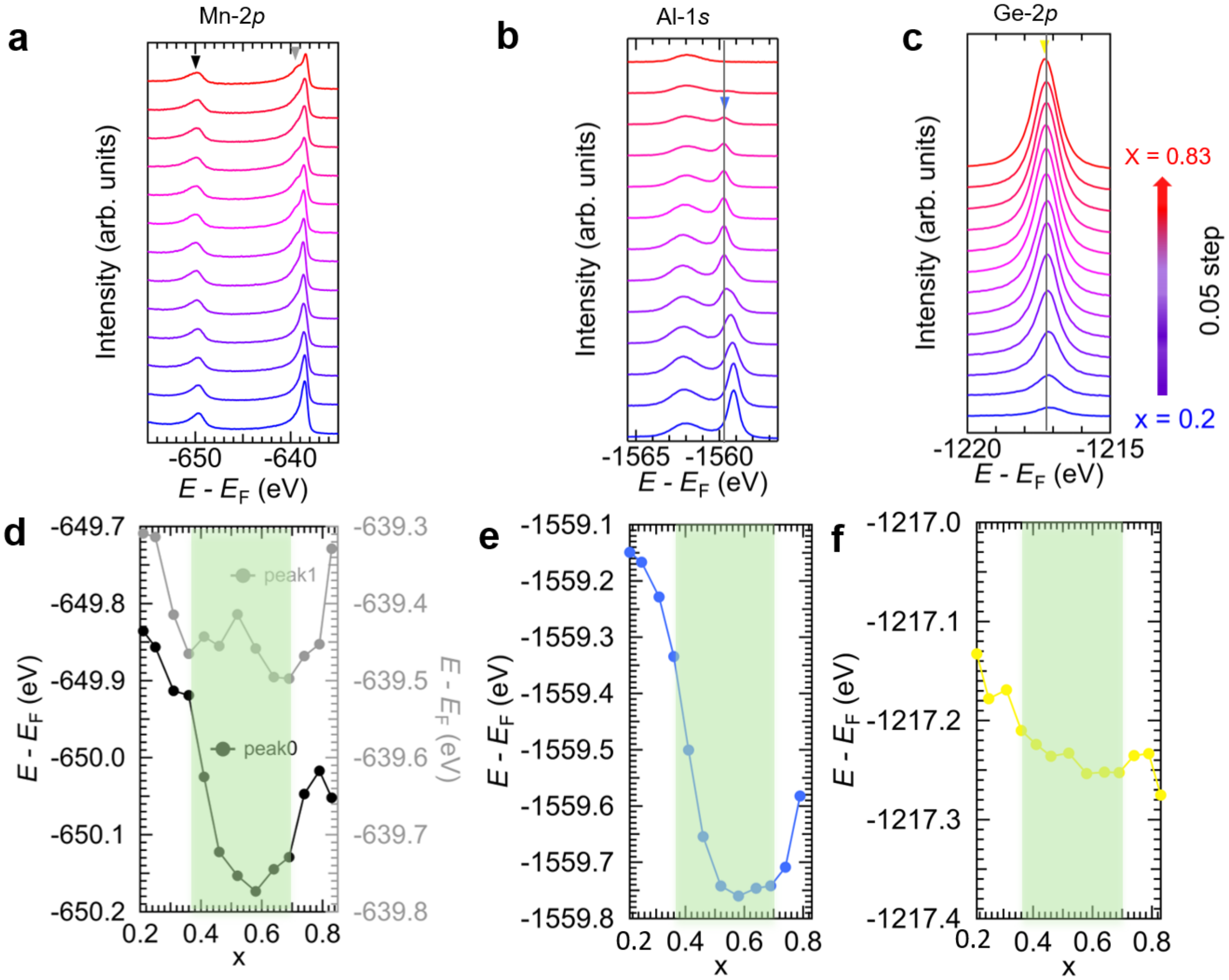

**Fig. 2**. Core-level spectroscopy of $Mn(Al_{1-x}Ge_x)_2$. a–c, HAXPES spectra of Mn 2*p*, Al 1*s*, and Ge 2*p* core levels for $0.2 \leq x \leq 0.83$. The spectra were analyzed using a photoemission fitting macro after subtracting a Shirley-type background, arrows and vertical guides indicate peak positions. d-f, Corresponding binding-energy evolution of Mn 2*p*, Al 1*s*, and Ge 2*p* as a function of Ge concentration, respectively.

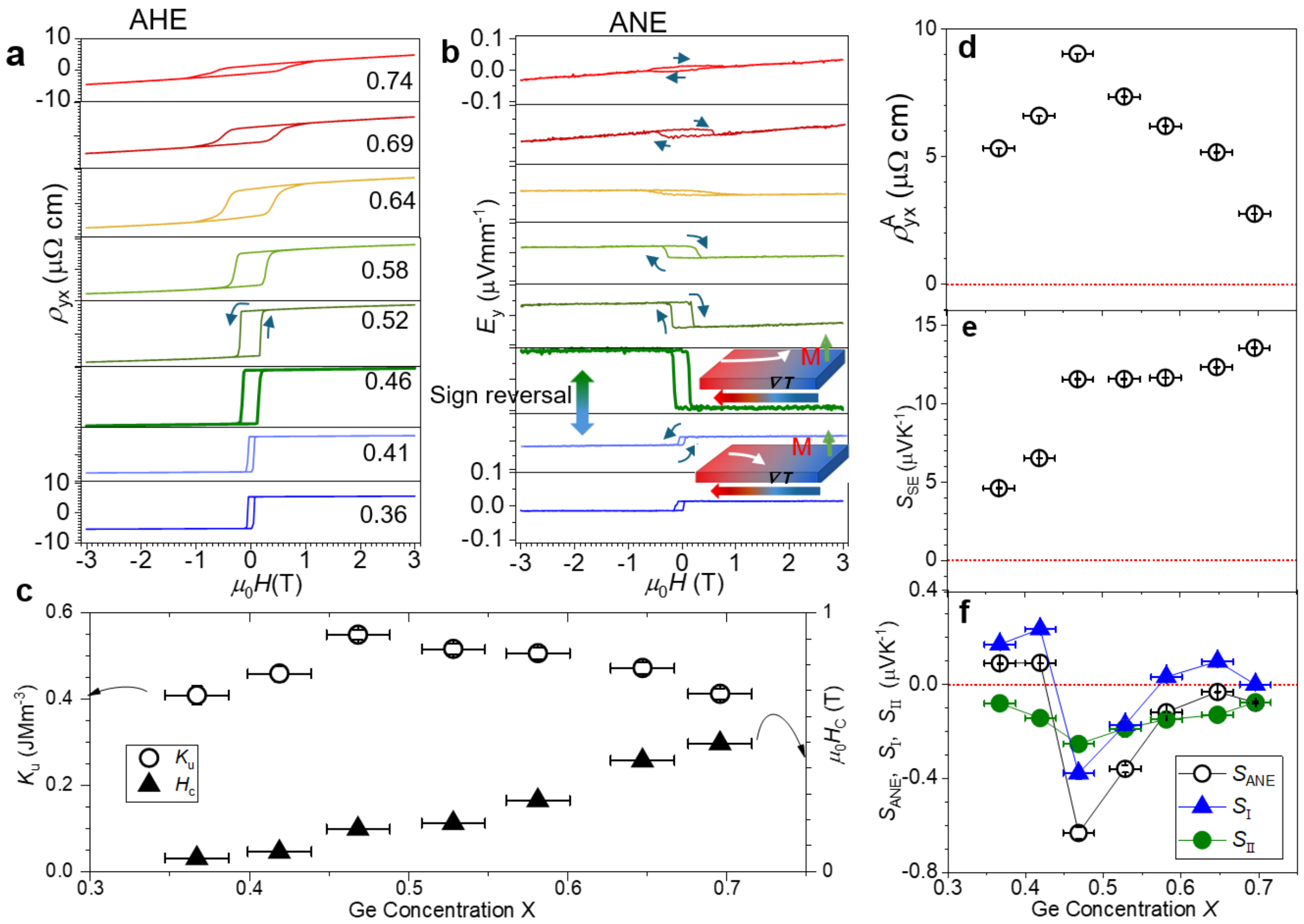

**Fig. 3**. Composition-dependent AHE and ANE in $Mn(Al_{1-x}Ge_x)_2$ thin films. a, $H$ dependence of $\rho_{yx}$ .(b) $H$ dependence of ANE-induced electric field $E_y$ at Ge concentration, $x$. Heater current was fixed at 450 mA for all measurements. The numbers in a, b, denote the Ge concentration, $x$. For clarity, repeated y-axis tick labels for $\rho_{yx}$ and $E_y$ are omitted, vertical scale bars indicate their respective magnitudes. c-e,e Ge concentration, $x$ dependence of $\rho^{\mathrm{A}}_{yx}$, $S_{\mathrm{SE}}$ $H_{\mathrm{c}}$, $H^{\mathrm{eff}}_{\mathrm{k}}$, $K_u$ and $S_{\mathrm{ANE}}$, and competing contributions $S_{\mathrm{I}}$ and $S_{\mathrm{II}}$, respectively.

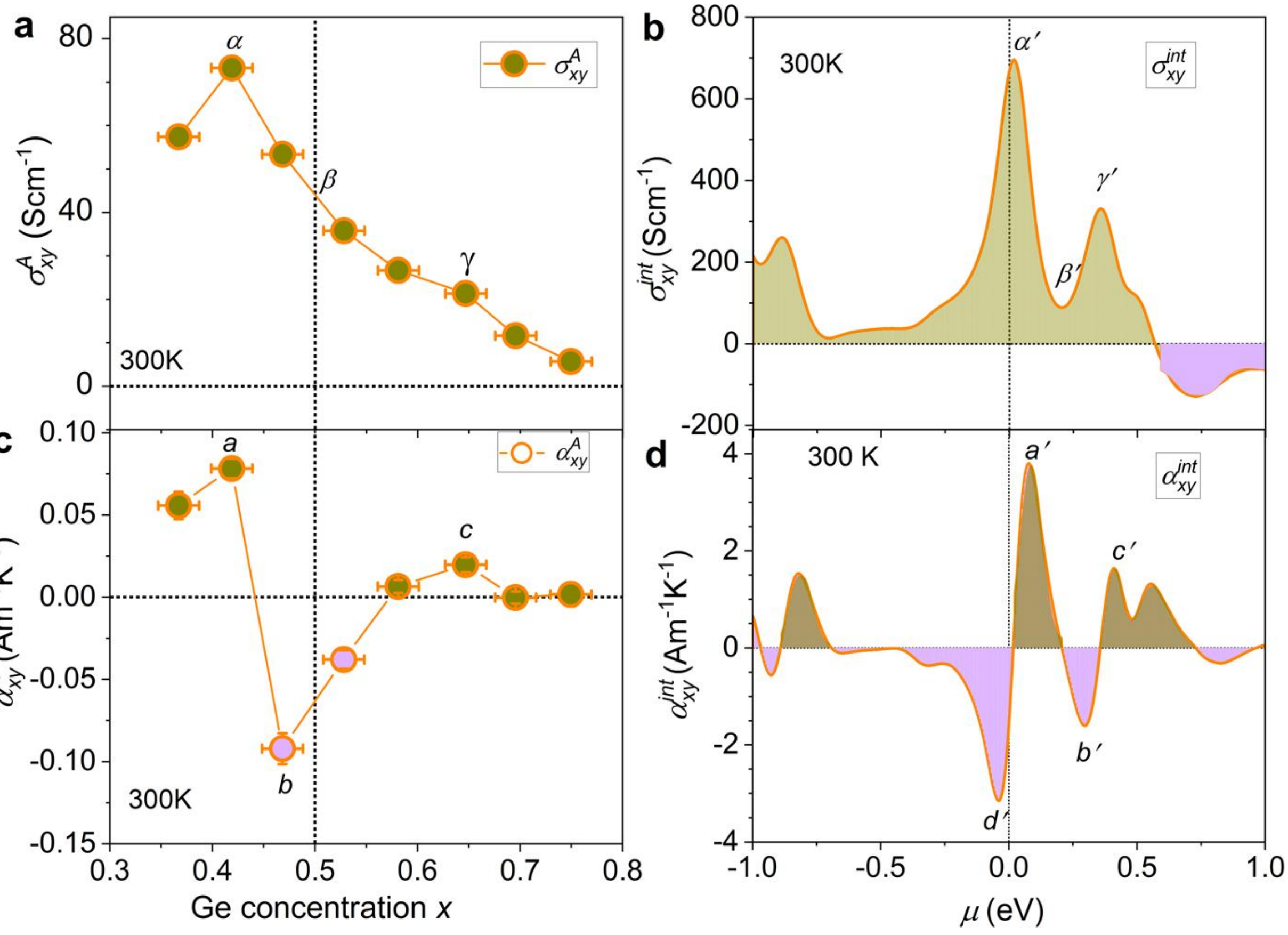

**Fig. 4.** Anomalous transport properties of $Mn(Al_{1-x}Ge_x)_2$. a,c, Experimental Ge concentration $x$ dependence of $\sigma_{xy}^{A}$ and $\alpha_{xy}^{A}$ evaluated in $Mn(Al_{1-x}Ge_x)_2$ composition spread film. b, d, Corresponding first-principles calculations of intrinsic $\sigma_{xy}^{int}$ and $\alpha_{xy}^{int}$ at 300 K, respectively.

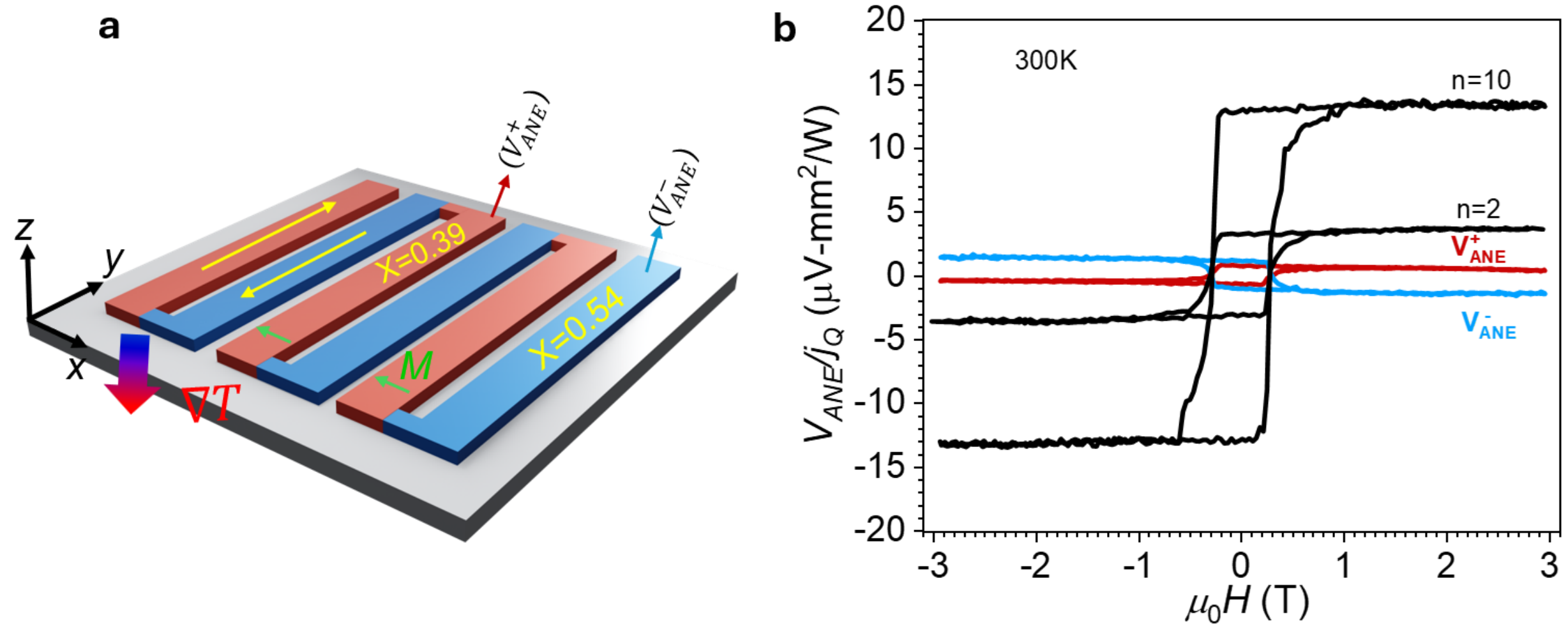

**Fig. 5.** Device demonstration of an ANE thermopile based on $Mn(Al_{1-x}Ge_x)_2$.a, Schematic illustration of the meander-type ANE thermopile structure, where positive (red) and negative (blue) $S_{ANE}$ elements are connected in series. b, ANE voltage response for devices with different numbers of positive/negative pairs, showing systematic scaling with device size.